\documentclass[superscriptaddress,reprint,prl]{revtex4-2}
\usepackage{amsmath,amssymb,amsthm}
\usepackage{newtxtext,newtxmath}
\usepackage{graphicx}
\usepackage{hyperref}
\usepackage{siunitx}
\usepackage{bm}
\usepackage{amsthm}
\usepackage{xcolor}
\definecolor{darkblue}{rgb}{0.0,0.0,0.5}
\hypersetup{colorlinks=true,linkcolor=darkblue,citecolor=darkblue,urlcolor=darkblue}
\usepackage{xurl}

\begin{document}
\title{Granularity Noise Limit in Atomic-Ensemble-Based Metrology}

\author{Chen-Rong Liu}
\thanks{These authors contributed equally to this work.}
\affiliation{College of Metrology Measurement and Instrument, China Jiliang University, Hangzhou, 310018 China}

\author{Chuang Li}
\thanks{These authors contributed equally to this work.}
\affiliation{College of Metrology Measurement and Instrument, China Jiliang University, Hangzhou, 310018 China}

\author{Runxia Tao}
\affiliation{College of Metrology Measurement and Instrument, China Jiliang University, Hangzhou, 310018 China}

\author{Yixuan Wang}
\affiliation{College of Metrology Measurement and Instrument, China Jiliang University, Hangzhou, 310018 China}

\author{Mingti Zhou}
\email{mtchou@cjlu.edu.cn}
\affiliation{College of Metrology Measurement and Instrument, China Jiliang University, Hangzhou, 310018 China}

\author{Xinqing Wang}
\email{wxqnano@cjlu.edu.cn}
\affiliation{College of Metrology Measurement and Instrument, China Jiliang University, Hangzhou, 310018 China}

\author{Ying Dong}
\email{yingdong@cjlu.edu.cn}
\affiliation{College of Metrology Measurement and Instrument, China Jiliang University, Hangzhou, 310018 China}

\date{\today}

\begin{abstract}
Conventional noise analysis in atomic‑ensemble sensing assumes a continuous‑medium approximation, thereby treating the atomic system as a deterministic dielectric. Here, we demonstrate that this assumption breaks down due to the discrete, particulate nature of the ensemble, giving rise to an intrinsic ``atomic granularity noise'' (AGN) that fundamentally competes with the optical measurement noise (OMN, typically photon shot noise). By introducing a discrete-atom statistical framework, we derive a unified noise-scaling law governed by a single dimensionless resource ratio, $\mathcal{R} = \bar{N}_{\mathrm{ph}}/\bar{N}_{\mathrm{at}}$ (the photon-to-atom flux ratio). This law predicts a continuous crossover from an OMN-limited regime to an AGN-limited regime. Crucially, our results reveal a counter-intuitive constraint for sensor optimization: increasing optical probe power—standard practice to mitigate OMN—can paradoxically degrade sensitivity by driving the system into the AGN-dominated regime. Furthermore, we identify a critical resource threshold, $\mathcal{R}_{\mathrm{crit}}$, beyond which quantum-enhanced metrology using non-classical light fails to improve sensitivity, as it becomes limited by the AGN.
\end{abstract}

\maketitle

\emph{Introduction.}—The predictive power of macroscopic field theories, from fluid dynamics to electrodynamics, relies on the continuous-medium hypothesis, which smooths the discrete nature of microscopic constituents into deterministic bulk parameters. While successful in the thermodynamic limit, this approximation breaks down when system sizes decrease or the discrete nature of matter can no longer be ignored. In fluid dynamics, for instance, the Navier-Stokes equations fail to describe strong gradients in shock waves~\cite{mott1951solution} or flows near moving contact lines~\cite{koplik1997molecular}. A similar breakdown occurs in nanoscale electromagnetism, where the standard continuous description of interfaces using uniform bulk permittivity becomes inadequate at the atomic scale~\cite{yang2019general}. Likewise, in nanofluidics, discrete atomic structure leads to flow enhancements that defy classical no-slip boundary conditions~\cite{majumder2005nanoscale}, while in granular matter, local rheology yields to non-local constitutive relations governed by grain discreteness~\cite{kamrin2012nonlocal, kim2020power}. Furthermore, even in small stochastic systems, large fluctuations can lead to transient violations of macroscopic thermodynamic laws~\cite{wang2002experimental}.

\begin{figure}[ht]
\centering
\includegraphics[width=\linewidth]{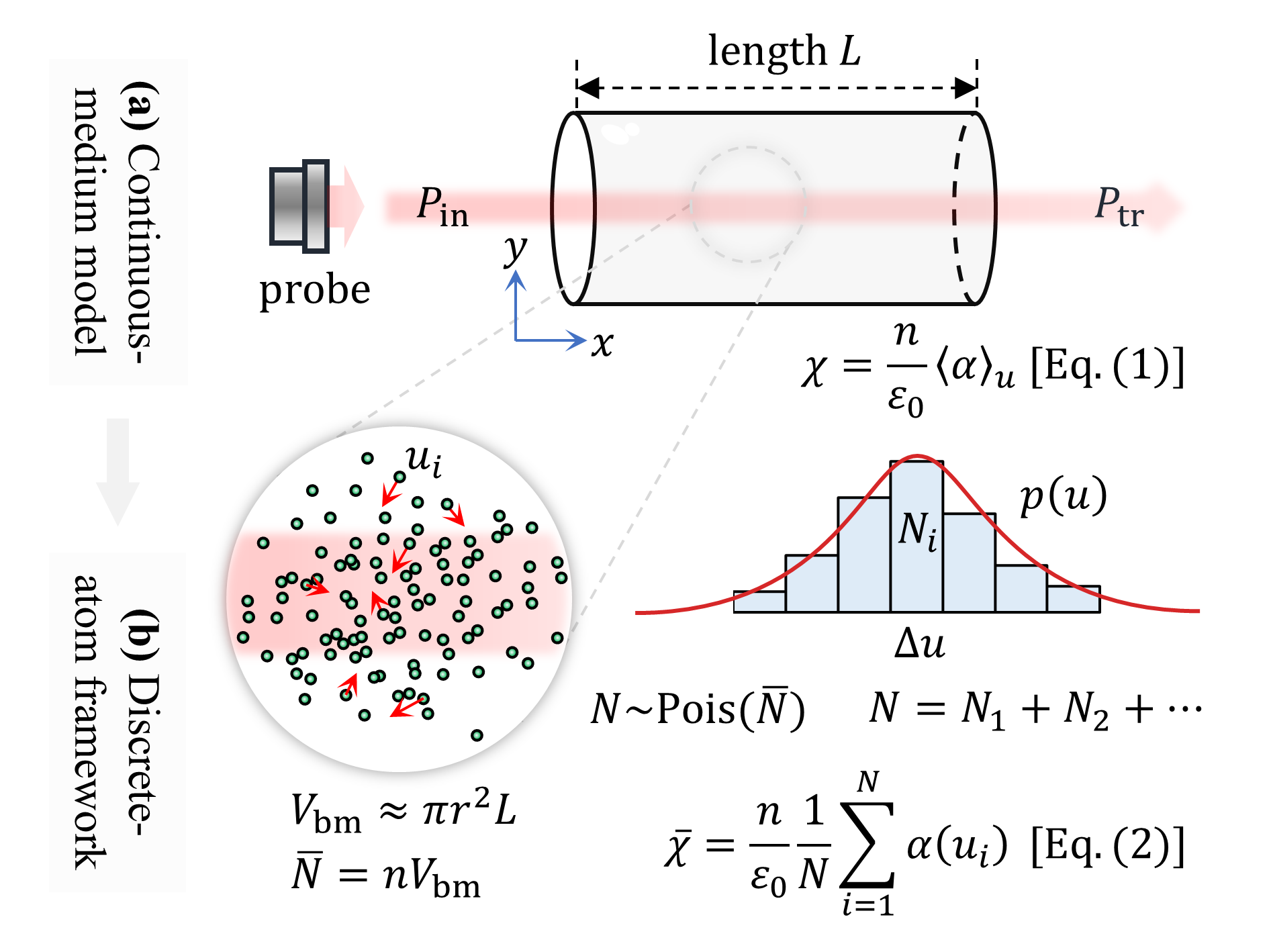}
\caption{(Color online) \textbf{Continuum versus discrete frameworks.} (a) The conventional continuous-medium model yields a deterministic susceptibility $\chi$ from a fixed atomic density $n$. (b) Our discrete-atom framework: atomic transit leads to a stochastic and finite number of atoms $N$ within the probe volume, and each atom possesses a random polarizability $\alpha_i$ that depends on a stochastic parameter $u_i$ (e.g., velocity or spin orientation). The resulting empirical distribution (histogram) exhibits finite-sample fluctuations around the underlying distribution $p(u)$, rendering the measured susceptibility $\bar{\chi}$ inherently stochastic.}
\label{fig:model}
\end{figure}

Here, we identify and formalize an analogous breakdown of the continuous-medium model in atomic-ensemble-based quantum metrology. Conventionally, vapor cells used in atomic sensing—ranging from optical magnetometers~\cite{Kominis2003, Budker2007} to Rydberg electrometers~\cite{Sedlacek2012, Adams2020}—are modeled as homogeneous dielectrics with deterministic susceptibility (cf. Fig.~\ref{fig:model}a). Consequently, their ultimate sensitivity is often assumed to be limited exclusively by photon shot noise (PSN)~\cite{wolfgramm2010squeezed,Jing2020}. Yet, this mean-field description overlooks a fundamental physical reality: the sensor consists of a finite, stochastic number of discrete atoms (see Fig.~\ref{fig:model}b). This atomic discreteness manifests as \emph{atomic granularity noise } (AGN)—intrinsic fluctuations in susceptibility originating from the finite sampling of random microscopic variables, such as thermal velocities~\cite{sagle1996measurement} or spin orientations~\cite{wolfgramm2010squeezed}. This discrete, stochastic picture—akin to the role of finite particle numbers in dissolving optical bistability cycles in cavity QED~\cite{rempe1991optical}, limit spin noise spectroscopy~\cite{crooker2004spectroscopy}, or set the projection noise limit in timekeeping~\cite{itano1993quantum}—implies that atomic granularity can impose a universal sensitivity limit competing directly with PSN.

In this Letter, we introduce a discrete‑atom statistical framework that bridges the regime limited by optical measurement noise (OMN, typically PSN) and that limited by atomic granularity. We derive a unified scaling law governed by a single dimensionless resource ratio $\mathcal{R}$ (the photon-to-atom flux ratio). This law dictates a continuous crossover between two fundamental regimes: (i) the OMN‑limited regime at small $\mathcal{R}$, where sensitivity is bounded by OMN; and (ii) the AGN‑limited regime at large $\mathcal{R}$, where the noise floor is dominated by AGN. Using Rydberg electrometry as a canonical testbed~\cite{Jing2020}, we uncover a counter-intuitive consequence for sensor optimization: increasing optical probe power does not indefinitely improve signal-to-noise ratio (SNR); instead, it eventually drives the sensor into a regime dominated by AGN, thereby degrading sensitivity. Crucially, our analysis identifies a hard boundary for quantum-enhanced metrology: a critical resource ratio, $\mathcal{R}_{\mathrm{crit}}$, beyond which atomic granularity prevails, rendering non-classical light resources ineffective. Beyond this specific platform, this framework is general and directly applicable to other prominent sensors such as optical magnetometers based on polarization rotation.

\emph{General theoretical framework.}—To elucidate the general framework, we begin with the paradigmatic system: a dilute, homogeneous atomic vapor. The susceptibility in the continuous-medium limit~\cite{fleischhauer2005electromagnetically} is
\begin{equation}\label{eq:chi_cont}
\chi = \frac{n}{\epsilon_0} \langle \alpha \rangle_u,
\end{equation}
where $n$ is the atomic density, $\epsilon_0$ is the vacuum permittivity, and $\langle \alpha \rangle_u$ is the expected single-atom polarizability $\alpha=\alpha_R+i\alpha_I$ over the stochastic parameter $u$ (e.g., velocity or spin orientation) with distribution $p(u)$.

This breaks down for a finite probe volume $V_{\mathrm{bm}}$ because the number of atoms $N$ within this volume is itself finite and stochastic. Specifically, $N$ follows Poisson statistics with mean $\bar{N}_{\mathrm{at}} = n V_{\mathrm{bm}}$, reflecting the random transit of atoms through the probe beam  region~\cite{sagle1996measurement}. Moreover, each atom contributes a polarizability $\alpha(u_i)$, where $u_i$ is an independent draw from $p(u)$. Consequently, the measured susceptibility becomes a stochastic \emph{sample mean}:
\begin{equation}\label{eq:chi_disc}
\bar{\chi} = \frac{n}{\epsilon_0}\cdot\frac{1}{N}\sum_{i=1}^{N} \alpha(u_i) \equiv \bar{\chi}_R + i\bar{\chi}_I .
\end{equation}
The expectation of this sample mean recovers the susceptibility [Eq.~(\ref{eq:chi_cont})], with $\langle N^{-1}\sum_{i} \alpha(u_i) \rangle = \langle \alpha \rangle_u$; however, the fluctuations of $\bar{\chi}$ constitute the intrinsic AGN. This discrete-atom statistical picture forms the foundation of our framework. 

The statistical properties of $\bar{\chi}$ follow from the independence of the atoms. For $\bar{N}_{\mathrm{at}} \gg 1$, the central limit theorem implies that $\bar{\chi}$ is approximately bivariate Gaussian; consequently, the variance of a quadrature component $q \in \{R, I\}$ is given by:
\begin{equation}\label{eq:Vq}
\mathrm{Var}(\bar{\chi}_q) \approx \frac{1}{\bar{N}_{\mathrm{at}}} \underbrace{\frac{n^2}{\epsilon_0^2} \mathrm{Var}_{u}[\alpha_q]}_{\mathcal{V}_q}.
\end{equation}
The quantity $\mathcal{V}_q \equiv (n^2/\epsilon_0^2) \mathrm{Var}_{u}[\alpha_q]$ defines the \emph{intrinsic variance}, which captures the quadrature fluctuations inherent to the microscopic distribution $p(u)$. (A detailed derivation is provided in Supplemental Material~\cite{supplemental_material}, Sec.~S1.)

In a typical optical readout, the extracted variable depends on both the atomic susceptibility and the photon count $\xi$: $\mathcal{S}=f(\bar{\chi}_q,\xi)$. Linearizing around the mean operating point $X_0\equiv(\langle\bar{\chi}_q\rangle,\bar{N}_{\mathrm{ph}})$ yields:
\begin{equation}\label{eq:signalModel}
\mathcal{S} = \mathcal{S}_0 + \mathcal{G} \, \delta\bar{\chi}_q + \mathcal{E},
\end{equation}
where $\mathcal{S}_0=f(X_0)$ is the mean signal, $\mathcal{G}\equiv\left.\partial f/\partial \bar{\chi}_q\right|_{X_0}$ is the transduction coefficient for atomic fluctuations $\delta\bar{\chi}_q$, and $\mathcal{E}$ represents the OMN arising from photon-counting statistics. For the canonical case of coherent light, where $\xi$ obeys Poisson statistics ($\mathrm{Var}(\xi)=\bar{N}_{\mathrm{ph}}$), the linear expansion yields $\sigma_{\mathcal{E}}^2 = \nu^2/\bar{N}_{\mathrm{ph}}$, governed by the photon transduction coefficient $\nu \equiv \left.\partial f/\partial(\xi/\bar{N}_{\mathrm{ph}})\right|_{X_0}$. Since the atomic fluctuations $\delta\bar{\chi}_q$ and the OMN $\mathcal{E}$ are statistically independent, combining Eqs.~(\ref{eq:Vq}) and (\ref{eq:signalModel}) yields the actual signal variance:
\begin{equation}\label{eq:actual_variance}
\sigma_{\mathcal{S}}^2 = \mathcal{G}^2\,\mathrm{Var}(\bar{\chi}_q) + \sigma_{\mathcal{E}}^2 = \frac{\mathcal{G}^2 \mathcal{V}_q}{\bar{N}_{\mathrm{at}}} + \frac{\nu^2}{\bar{N}_{\mathrm{ph}}}.
\end{equation}
The competition is now explicit: AGN scales as $1/\bar{N}_{\mathrm{at}}$, while OMN scales as $1/\bar{N}_{\mathrm{ph}}$. This competition is governed by two dimensionless parameters: the \emph{resource ratio} $\mathcal{R} \equiv \bar{N}_{\mathrm{ph}}/\bar{N}_{\mathrm{at}}$ and the \emph{intrinsic fluctuation parameter} $\mathcal{J} \equiv \mathcal{G}^2 \mathcal{V}_q / \nu^2$. Normalizing the actual noise by the OMN limit, $\sigma_{\mathcal{E}}$, yields the unified scaling law:
\begin{equation}\label{eq:scaling_law}
\frac{\sigma_{\mathcal{S}}}{\sigma_{\mathcal{E}}} = \sqrt{1 + \mathcal{R} \, \mathcal{J}}.
\end{equation}

Equation~(\ref{eq:scaling_law}) encapsulates the universal resource competition, describing a continuous crossover between two fundamental regimes separated by the critical threshold $\mathcal{R}_c \equiv 1/\mathcal{J}$: (i)~the OMN‑limited regime ($\mathcal{R} \ll \mathcal{R}_c$), where $\sigma_{\mathcal{S}} \approx \sigma_{\mathcal{E}} \propto 1/\sqrt{\bar{N}_{\mathrm{ph}}}$ and sensitivity is bounded by photon‑counting statistics; and (ii)~the AGN‑limited regime ($\mathcal{R} \gg \mathcal{R}_c$), where $\sigma_{\mathcal{S}} \approx \sqrt{\mathcal{R}\mathcal{J}} \, \sigma_{\mathcal{E}} \propto 1/\sqrt{\bar{N}_{\mathrm{at}}}$ and sensitivity is fundamentally constrained by AGN. The crossover from the conventional shot-noise-dominated (OMN‑limited) regime to the novel AGN‑limited regime is thus determined by whether the resource ratio $\mathcal{R}$ exceeds this intrinsic threshold.

\begin{figure*}[ht]
\centering
\includegraphics[width=\linewidth]{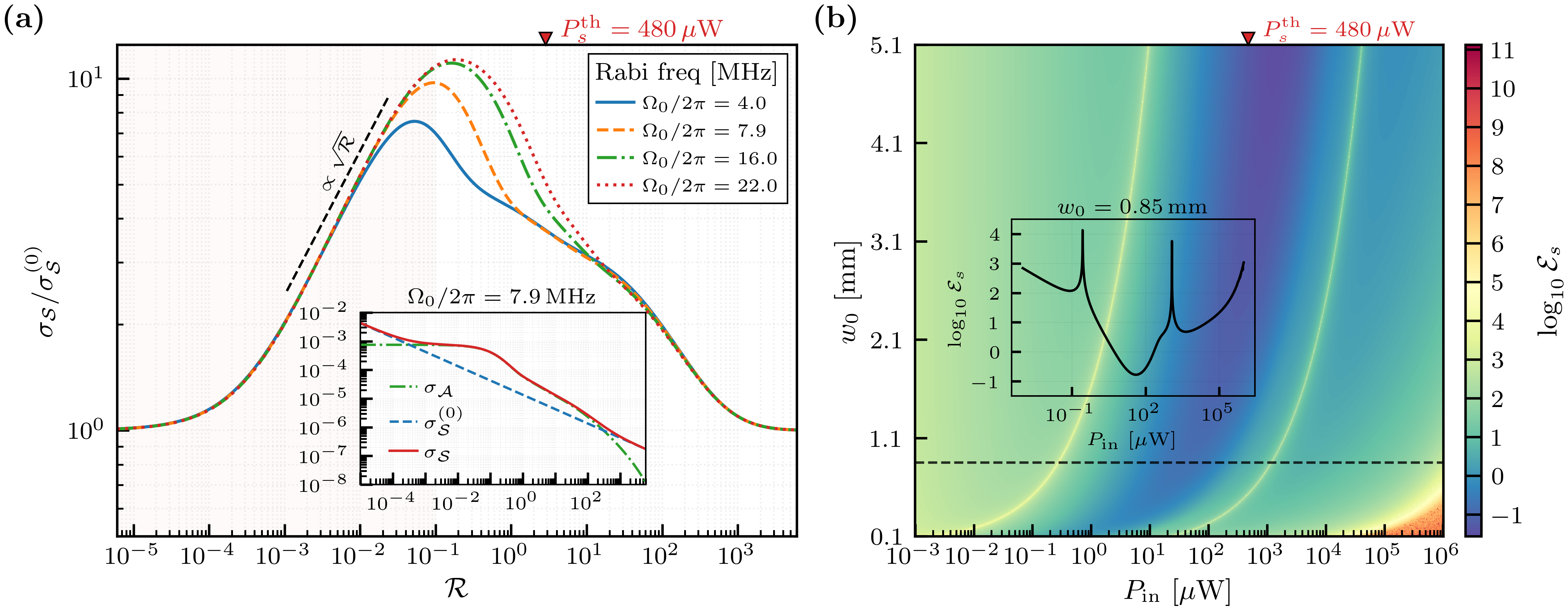}
\caption{(Color online) \textbf{Noise scaling and sensitivity.}
(a) Main panel: relative noise $\sigma_{\mathcal{S}}/\sigma_{\mathcal{S}}^{(0)}$ [Eq.~(\ref{eq:scaling_Ryd})] vs resource ratio $\mathcal{R}$ for several Rabi frequencies $\Omega_0$, with all three transitions on resonance. Under the weak-probe condition ($P\ll P_s^{\mathrm{th}}=480\,\mu$W, the Doppler-corrected saturation power), the curves exhibit a continuous crossover from the OMN-limited regime ($\mathcal{R}\ll\mathcal{R}_c$) to the AGN-limited regime ($\mathcal{R}\gg\mathcal{R}_c$), delineated by the critical threshold $\mathcal{R}_c\equiv1/\mathcal{J}$; in the AGN-limited regime the noise scales as $\sqrt{\mathcal{R}}$. 
Inset: for $\Omega_0/2\pi=7.9$~MHz, the total noise $\sigma_{\mathcal{S}}$, the AGN contribution $\sigma_{\mathcal{A}}\equiv\sqrt{\mathcal{J}/\bar{N}_{\mathrm{at}}}$, and the PSN limit $\sigma_{\mathcal{S}}^{(0)}=1/\bar{N}_{\mathrm{ph}}$ vs $\mathcal{R}$, showing AGN dominating once $\mathcal{R}$ exceeds $\mathcal{R}_c$.
(b) Microwave electric-field measurement sensitivity $\mathcal{E}_s$ [Eq.~(\ref{eq:sensitivity})] vs input power $P_{\mathrm{in}}$ and beam waist $w_0$, displayed as a $\log_{10}\mathcal{E}_s$ heatmap. The inset shows $\mathcal{E}_s$ vs $P_{\mathrm{in}}$ at a fixed beam waist $w_0=0.85$~mm (dashed line), corresponding to the experimental parameters of Refs.~\onlinecite{Jing2020, Liu2026Fisher}. The results are based on the four-level Rydberg electrometry scheme~\cite{Jing2020, Liu2026Fisher}.}
\label{fig:noiseRatio}
\end{figure*}

\emph{Application to Rydberg electrometry.}—We now exemplify the general framework using the canonical case of Rydberg electrometry. In optical transmission measurements, the recorded photon count $\xi$ over a time interval $\Delta t$ follows a Poisson distribution with mean $\bar{N}_{\mathrm{ph}} = P_{\mathrm{in}}\Delta t / (\hbar\omega_p)$, where $P_{\mathrm{in}}$ is the incident power and $\omega_p$ the optical angular frequency~\cite{steckQuantumAtomOptics}. The stochastic optical depth for a probe laser of wavelength $\lambda_p$ traversing a vapor cell of length $L$ is $a\bar{\chi}_I$, with $a \equiv 2\pi L/\lambda_p$ and $\bar{\chi}_I$ the imaginary part of the susceptibility defined in Eq.~(\ref{eq:chi_disc}). The extracted optical depth $\mathcal{S} \equiv \ln(P_\mathrm{in}/P)$ becomes
\begin{equation}\label{eq:S_def_app}
\mathcal{S} = a\bar{\chi}_I - \ln\left(\xi/\bar{N}_{\mathrm{ph}}\right).
\end{equation}
Considering $\bar{N}_{\mathrm{ph}} \gg 1$, the linearization $\ln(\xi/\bar{N}_{\mathrm{ph}}) \approx (\xi - \bar{N}_{\mathrm{ph}})/\bar{N}_{\mathrm{ph}}$ is justified. Hence, Eq.~(\ref{eq:S_def_app}) maps directly onto the linear-response form of Eq.~(\ref{eq:signalModel}) with the identifications $q=I$, $\mathcal{S}_0 = a\langle\bar{\chi}_I\rangle$, $\mathcal{G} = a$, and $\mathcal{E} = (\bar{N}_{\mathrm{ph}}-\xi)/\bar{N}_{\mathrm{ph}}$ (specifically PSN). A detailed derivation is provided in Supplemental Material Sec.~S2 (see also Refs.~\cite{Tanasittikosol2011,omont1977collisions,fabrikant1986interaction,Fan2015Microwave,kumar2017atom,otto2022towards,fontaine2020,miller2024rydiqule,nagib2025fast} therein). Substituting these into Eq.~(\ref{eq:actual_variance}), the variance of the optical depth becomes:
\begin{equation}\label{eq:varSryd}
\sigma_{\mathcal{S}}^2 =  a^2\frac{\mathcal{V}_I}{\bar{N}_{\mathrm{at}}} + \frac{1}{\bar{N}_{\mathrm{ph}}},
\end{equation}
where $\mathcal{V}_I = (n^2/\epsilon_0^2)\mathrm{Var}_{u}[\alpha_I]$ is the intrinsic variance, dominated here by the Maxwell‑Boltzmann velocity distribution. In this system, the intrinsic fluctuation parameter reduces to $\mathcal{J} = a^2\mathcal{V}_I$, and $\sigma_{\mathcal{A}} \equiv \sqrt{\mathcal{J}/\bar{N}_{\mathrm{at}}}$ is the AGN contribution. Together with the resource ratio $\mathcal{R} \equiv \bar{N}_{\mathrm{ph}}/\bar{N}_{\mathrm{at}}$, we thus obtain a specific instance of the unified scaling law [Eq.~(\ref{eq:scaling_law})]:
\begin{equation}\label{eq:scaling_Ryd}
\frac{\sigma_{\mathcal{S}}}{\sigma_{\mathcal{S}}^{(0)}} = \sqrt{1 + \mathcal{R} \, \mathcal{J}},
\end{equation}
where $\sigma_{\mathcal{S}}^{(0)} = 1/\sqrt{\bar{N}_{\mathrm{ph}}}$ is the shot-noise limit (i.e., OMN).

Figure~\ref{fig:noiseRatio}(a) plots the relative noise $\sigma_{\mathcal{S}}/\sigma_{\mathcal{S}}^{(0)}$ versus $\mathcal{R}$ for several representative Rabi frequencies $\Omega_0/2\pi$ (see legend). Under the weak-probe condition ($P \ll P_s^{\mathrm{th}}$, where $P_s^{\mathrm{th}}$ is the Doppler-corrected saturation power~\cite{steckQuantumAtomOptics}; marked by red $\blacktriangledown$ in the panel), the curves exhibit the predicted continuous crossover ($\mathcal{R}_c\sim 10^{-3}$) from the OMN-limited to the AGN-limited regime. In the OMN-limited regime ($\mathcal{R} \ll \mathcal{R}_c$), the noise converges to unity (the shot-noise limit). In the AGN-limited regime ($\mathcal{R} \gg \mathcal{R}_c$), the noise scales as $\sqrt{\mathcal{R}}$, and is largely insensitive to $\Omega_0$. However, at very large $\mathcal{R}$, the scaling breaks down as $\mathcal{J}(\mathcal{R})$ itself becomes dependent on $\mathcal{R}$—a consequence of the intensity-dependent atomic polarizability beyond the weak-probe regime.

The universal scaling law provides a concrete criterion to assess AGN dominance. Physically, the resource ratio $\mathcal{R}$ directly compares the incident photon flux $\Phi_{\mathrm{ph}}=P_{\mathrm{in}}/(\hbar \omega_p)$ to the atomic refresh flux $\Phi_{\mathrm{at}}$ within the probe volume. We estimate $\Phi_{\mathrm{at}}$ by modeling the beam as a cylinder of length $L$ and radius $w_{p0}$ (volume $V_{\mathrm{bm}} = \pi w_{p0}^{2} L$). In steady state, kinetic theory gives the atomic flux through the sidewall area $A=2\pi w_{p0} L$ as $\Phi_{\mathrm{at}} = n\,\bar{v}\,A/4$, where $\bar{v}=\sqrt{8k_BT/\pi m}$ is the mean thermal speed for atoms of mass $m$ at temperature $T$. For parameters representative of a contemporary Rydberg electrometry experiment (see Ref.~\cite{Jing2020} and Supplemental Material Table~S2.1)---a room-temperature $^{133}$Cs vapor cell ($n \approx 4.89 \times 10^{10}\,\mathrm{cm}^{-3}$, $T=298.15\,\mathrm{K}$) probed by a focused beam ($w_{p0}=0.85\,\mathrm{mm}$, $L=5\,\mathrm{cm}$) with $P_{\mathrm{in}}=120\,\mu\mathrm{W}$ at $852\,\mathrm{nm}$, i.e., one quarter of the Doppler-corrected saturation power $P_s^{\mathrm{th}}=480\,\mu\mathrm{W}$ and thus above the weak-probe regime (see Fig.~\ref{fig:noiseRatio}(a))---we obtain $\Phi_{\mathrm{at}} \approx 7.1 \times 10^{14}\,\mathrm{s}^{-1}$ and $\Phi_{\mathrm{ph}} \approx 5.1 \times 10^{14}\,\mathrm{s}^{-1}$, yielding an experimental resource ratio $\mathcal{R} = \Phi_{\mathrm{ph}}/\Phi_{\mathrm{at}} \approx 0.72$. Crucially, for moderate microwave driving ($\Omega_0/2\pi \sim 7.9,\mathrm{MHz}$~\cite{Jing2020}), our analysis yields $\mathcal{J}\approx 40$ (see Supplemental Material Fig.~S2.2), leading to a relative noise $\sigma_{\mathcal{S}}/\sigma_{\mathcal{S}}^{(0)}\approx 5.4$, placing the sensor in the AGN-limited regime, albeit near saturation.

Physically, the AGN-limited regime reflects a fundamental constraint: within the probe volume, the atomic refresh flux is insufficient to average out the granularity of discrete atoms before they are interrogated by the incident photon flux. This refresh-rate bottleneck, previously overlooked, sets a fundamental floor on sensitivity when $\mathcal{R} \gg \mathcal{R}_c$ under the weak-probe condition, where $\mathcal{J}$ remains constant and the noise scales as $\sqrt{\mathcal{R}}$. Beyond the weak-probe regime, however, the atomic polarizability becomes intensity-dependent, causing $\mathcal{J}(\mathcal{R})$ to acquire a convex dependence on $\mathcal{R}$. This modifies the noise scaling to $\sigma_{\mathcal{S}}/\sigma_{\mathcal{S}}^{(0)} = \sqrt{1+\mathcal{R}\,\mathcal{J}(\mathcal{R})}$, which exhibits a non-monotonic behavior—initially increasing with $\mathcal{R}$ and then decreasing after reaching a maximum. The analyzed experimental condition ($\mathcal{R}\approx0.72$) lies on the descending branch of this curve, placing the sensor in the AGN-limited regime but with a relative noise level ($\approx5.4$) that already falls below the extrapolated $\sqrt{\mathcal{R}}$ scaling from the deep AGN limit.

Building on this framework, when measuring a small microwave field change $\delta\Omega_s$ via slope detection, the measurement uncertainty obtained via error propagation is $\sigma_{\Omega_s} = \sigma_{\mathcal{S}}/|\partial\mathcal{S}_0/\partial\Omega_s|$. Setting SNR $=1$ then yields the fundamental sensitivity~\cite{degen2017quantum,Liu2026Fisher}:
\begin{equation}\label{eq:sensitivity}
\mathcal{E}_{s} = \frac{\hbar}{\mu_s}[\delta\Omega_s]_{\mathrm{SNR}=1}\cdot\sqrt{T} = \mathcal{E}_{s}^{(0)} \, \sqrt{1 + \mathcal{R}\,\mathcal{J}},
\end{equation}
where $\mathcal{E}_{s}^{(0)}$ is the corresponding shot-noise-limited sensitivity. Here, $\mu_s$ is the transition dipole moment of the microwave‑driven Rydberg transition, $T$ is the integration time, and $[\delta\Omega_s]_{\mathrm{SNR}=1}$ is the minimal detectable Rabi frequency. Under weak-probe condition, it follows that achieving the shot‑noise limit requires operating in the OMN‑limited regime ($\mathcal{R} \ll \mathcal{R}_c$), which necessitates minimizing the resource ratio $\mathcal{R}$. Counter‑intuitively, the conventional strategy of increasing optical power $P_{\mathrm{in}}$ to improve the SNR inherently raises $\mathcal{R} \propto P_{\mathrm{in}}$. Once $\mathcal{R} \gtrsim \mathcal{R}_c$, the system enters the AGN‑limited regime, where the concomitant rise in AGN (in this approximation scaling as $\sqrt{\mathcal{R}} \propto \sqrt{P_{\mathrm{in}}}$) eclipses the benefit of reduced photon noise ($\sigma_{\mathcal{E}} \propto 1/\sqrt{P_{\mathrm{in}}}$), thereby degrading the overall sensitivity. Consequently, optimal performance is not achieved by merely increasing optical power, but by balancing the photon and atom fluxes to keep $\mathcal{R}$ below the critical threshold $\mathcal{R}_c$. Indeed, although the experimental parameters of Ref.~\cite{Jing2020} place the operation above the strict weak-probe regime, its operating point in Fig.~\ref{fig:noiseRatio}(b) (inset) lies near the sensitivity optimum—a direct manifestation of the AGN–OMN trade-off captured by the scaling law.

\emph{Extension to quantum-enhanced sensing.}—The universal scaling law [Eq.~(\ref{eq:scaling_law})], with its explicit separation of atomic and photonic noise, naturally incorporates quantum-enhanced sensing by generalizing OMN. Specifically, we parametrize the photon-counting variance as $\mathrm{Var}(\xi) = \bar{N}_{\mathrm{ph}}(1+Q)$, where the Mandel parameter $Q \ge -1$ spans from Fock states ($Q=-1$) to coherent ($Q=0$) and super-Poissonian ($Q>0$) light~\cite{gerry2023introductory}. For a given readout transduction $\nu$, the OMN variance generalizes to $\sigma_{\mathcal{E}}^2(1+Q)$. This extra factor is absorbed into a renormalized intrinsic fluctuation parameter, $\mathcal{J}_{Q} \equiv \mathcal{J}/(1+Q)$, which effectively rescales the impact of atomic granularity relative to the reduced optical noise. The generalized scaling law for the probe state thus takes the form:
\begin{equation}\label{eq:scaling_Q}
\frac{\sigma_{\mathcal{S}}}{\sigma_{\mathcal{E}}^{(Q)}} = \sqrt{1 + \mathcal{R} \, \mathcal{J}_Q},
\end{equation}
where $\sigma_{\mathcal{E}}^{(Q)} = \sigma_{\mathcal{E}}\sqrt{1+Q}$ is the quantum noise limit,  reduced below the shot-noise limit for $Q<0$.

\begin{figure}[ht]
\centering
\includegraphics[width=\linewidth]{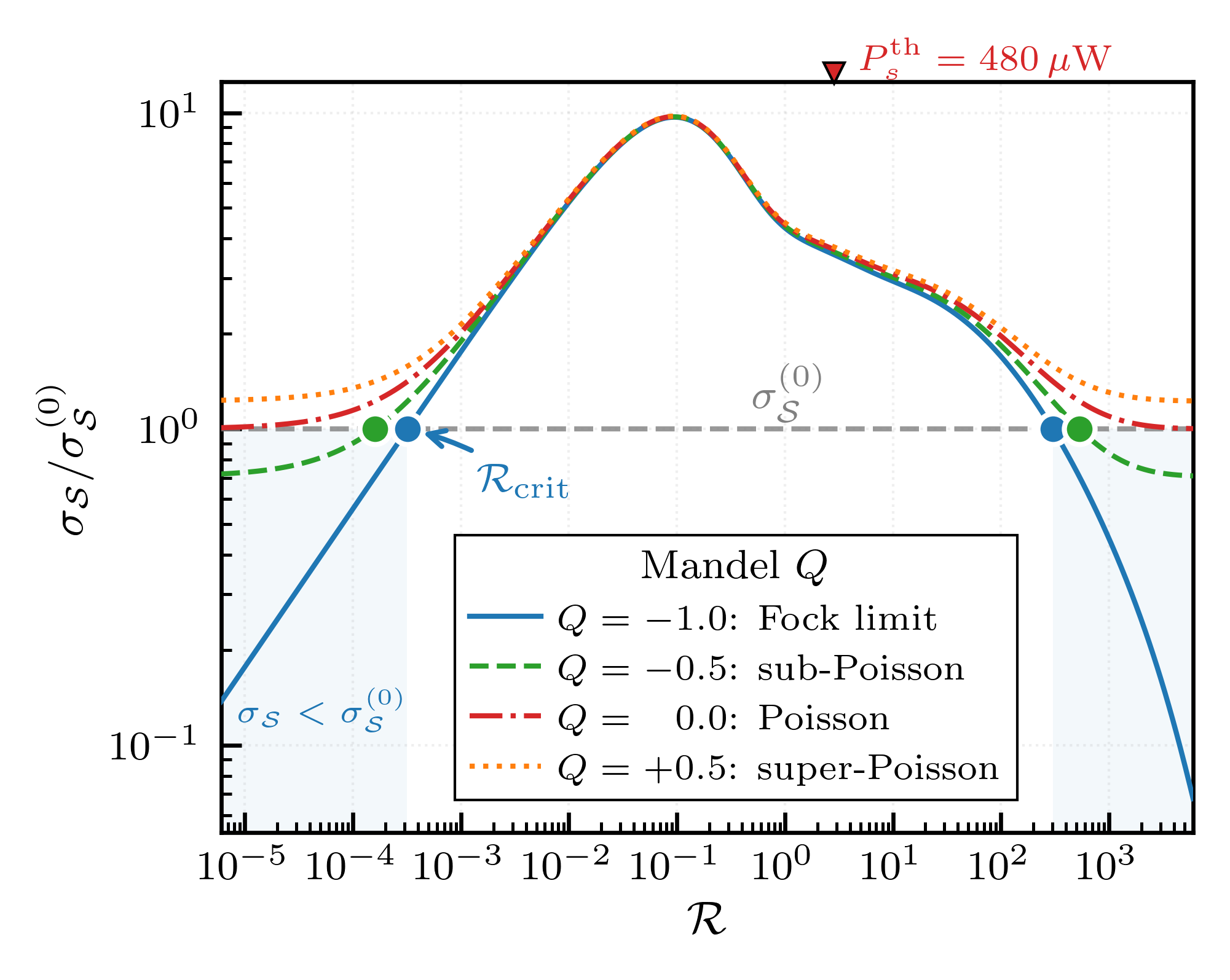}
\caption{(Color online) \textbf{Quantum enhancement limited by atomic granularity.} Relative noise $\sigma_{\mathcal{S}}/\sigma_{\mathcal{S}}^{(0)}$ versus resource ratio $\mathcal{R}$ for different photon statistics $Q$ (same axes as Fig.~\ref{fig:noiseRatio}; $\Omega_0/2\pi = 7.9\,\mathrm{MHz}$). While squeezing ($Q<0$) suppresses OMN, it exposes the AGN floor at a lower threshold. The critical ratio $\mathcal{R}_{\mathrm{crit}}$ (arrow) is defined by the intersection where the Fock-state limit ($Q=-1$, solid blue) meets the shot-noise limit (dashed horizontal line). For $\mathcal{R} > \mathcal{R}_{\mathrm{crit}}$, AGN prevails and strictly precludes any quantum advantage.}
\label{fig:noiseMandelRatio}
\end{figure}

Crucially, while sub-Poissonian light ($Q<0$) lowers the OMN floor, it paradoxically accelerates the onset of the AGN-limited regime by increasing the effective parameter $\mathcal{J}_Q > \mathcal{J}$, under weak-probe condition. Figure~\ref{fig:noiseMandelRatio} illustrates this competition using the same axes as Fig.~\ref{fig:noiseRatio}. We identify a fundamental \emph{quantum advantage boundary}, $\mathcal{R}_{\mathrm{crit}}$, defined physically by the condition $\sigma_{\mathcal{S}}|_{Q=-1} = \sigma_{\mathcal{S}}^{(0)}$. This point marks the hard threshold where the noise of the ideal Fock state (the ultimate quantum limit) matches the shot-noise limit. For $\mathcal{R} < \mathcal{R}_{\mathrm{crit}}$, quantum resources provide a sensitivity gain ($\sigma_{\mathcal{S}} < \sigma_{\mathcal{S}}^{(0)}$). However, once $\mathcal{R}$ exceeds $\mathcal{R}_{\mathrm{crit}}$, the sensor hits the ``granularity wall'' where AGN dominates even for the ideal Fock state ($Q=-1$), rendering non-classical light ineffective. Thus, $\mathcal{R}_{\mathrm{crit}}$ defines the maximal usable photon flux for quantum-enhanced metrology—a hard constraint imposed solely by the discrete nature of the atomic ensemble.

\emph{Conclusion.}—We have identified atomic granularity as an inescapable noise source, analogous to the breakdown of the continuum approximation observed in nanofluidics and granular matter. Our discrete-atom framework yields a unified scaling law [Eq.~(\ref{eq:scaling_law})] controlled by the resource ratio $\mathcal{R} = \bar{N}_{\mathrm{ph}}/\bar{N}_{\mathrm{at}}$, mapping the crossover from the conventional OMN‑limited regime to the newly identified AGN‑limited regime. This framework reveals a paradox in sensor optimization: increasing optical power eventually ceases to improve sensitivity as the system enters the AGN regime ($\mathcal{R} \gtrsim 1/\mathcal{J}$). Furthermore, it establishes a hard ``quantum-advantage horizon'' $\mathcal{R}_{\mathrm{crit}}$; beyond this point, atomic granularity dominates, rendering non-classical light ineffective. Optimal metrology thus requires balancing photonic and atomic fluxes to remain below these thresholds. Applicable to any sensor based on ensemble-averaged susceptibility, this work unifies the treatment of measurement noise and intrinsic fluctuations, providing a theoretical blueprint for the design of granularity‑aware atomic sensors.

\begin{acknowledgments}
This work was supported by the National Natural Science Foundation of China (Grant NO. 12304545 and 12475042) and the National Key Research and Development Program of China (Grant NO. 2023YFF0718400).
\end{acknowledgments}


\bibliography{StatTheoRydSensRef}

\end{document}